\documentclass{XrU2005}

\title{Dark Matter Halos of Early-Type Galaxies}
\author{P. J. Humphrey$^1$, D. A. Buote$^1$, F. Gastaldello$^1$, L. Zappacosta$^1$, J.~S. Bullock$^1$, F. Brighenti$^{2,3}$ \& W.~G. Mathews$^2$}
\affil{$^1$Dept. of Physics \& Astronomy, University of California, Irvine, Irvine, CA 92697\\$^2$University of California Observatories, Lick Observatory, University of California at Santa Cruz, Santa Cruz, CA 95064\\$^3$Dipartimento di Astronomia, Universit\`{a} di Bologna, Via Ranzani 1, Bologna 40127, Italy}

\usepackage{graphicx}
\usepackage{natbib}
\usepackage[hypertex]{hyperref}

\def \thin {\thinspace}
\def \nh {N${\rm _H}$}

\def\spose#1{\hbox to 0pt{#1\hss}}
\def\ltsim{$\mathrel{\spose{\lower 3pt\hbox{$\sim$}}
        \raise 2.0pt\hbox{$<$}}$\thinspace}
\def\gtsim{$\mathrel{\spose{\lower 3pt\hbox{$\sim$}}
        \raise 2.0pt\hbox{$>$}}$\thinspace}
\def \msun {${\rm M_\odot}$}
\def \lsun {${\rm L_\odot}$}

\def \nh {$N_{\rm H}$}

\def \eg {e.g.}

\def \dtwentyfive {${\rm D_{25}}$}
\newcommand{\reff}{${\rm R_{e}}$}

\newcommand{\mstars}{${\rm M_*}$}

\newcommand{\mvir}{${\rm M_{vir}}$}

\newcommand{\chandra }{{\em Chandra}}

\newcommand{\ciao }{{\em CIAO}}

\newcommand{\lb }{${\rm L_B}$}

\begin{document}


\maketitle

\begin{abstract}
Cosmological simulations of galaxy formation predict a universal form for
the mass profile of dark matter (DM) halos from cluster to galaxy scales.
Remarkably few interesting constraints exist, however, on DM halos in 
early-type galaxies. 
Using \chandra\ we present the temperature, density and
mass profiles of a small sample of early-type galaxies, revealing
significant DM in each case. When a component is 
included to account for stellar mass and the DM halo is allowed
to respond adiabatically to the baryonic condensation into stars,
the mass profiles are well-fitted by the universal profile, with
Virial masses and concentrations in agreement with simulations.
However, only $\sim$half, or less, of the mass within \reff\ seems
attributable to the stars, implying stellar \mstars/\lb\ $\sim$1--5.
\end{abstract}

\section{Introduction}
The nature of DM within the Universe is one of the 
fundamental problems facing modern physics. 
N-body cosmological simulations predict a 
``universal'' profile for DM halos over a wide range
of mass-scales \citep[hereafter NFW]{navarro97}. 
 In an hierarchical formation
scenario the early epoch of formation of low-mass halos 
should ``freeze in'' more tightly concentrated halos at the galaxy
scale than are observed in clusters \citep{bullock01a}. 
What is less clear, however, is the way in which the DM halo
responds to the condensation of baryons into stars.
If the 
galaxy is assembled by a series of mergers, however, the baryonic and 
dark matter may be mixed in such a way that the total gravitating
mass follows the NFW profile \citep{loeb03a}. Alternatively, 
present-day ellipticals may retain the ``memory'' of the 
original contraction \citep{gnedin04a}.
We present X-ray determined mass profiles for 7 early-type galaxies,
spanning the mass range $\sim 10^{12}$--$\sim 10^{13}$\msun,
chosen from the \chandra\ archive to be sufficiently
bright and relaxed  enough to yield interesting mass constraints.
Two companion posters, Gastaldello et al
and Zappacosta et al (both this volume) address DM halos at 
group and cluster scales.

\begin{figure}[!h]
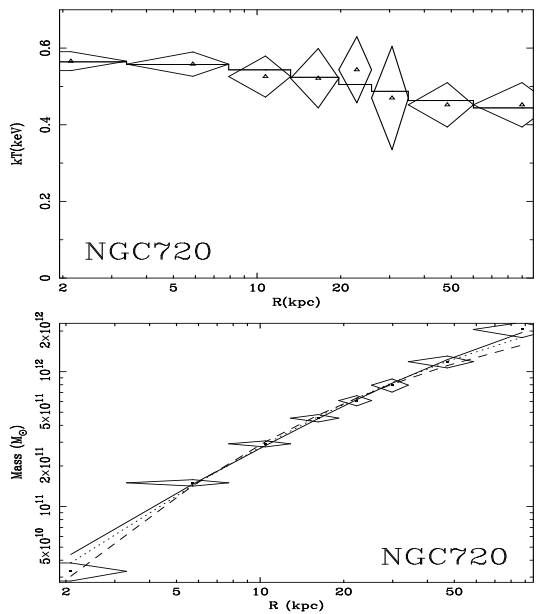

\centering
\includegraphics[height=7cm,width=4cm,angle=270]{fig1a.eps}
\includegraphics[height=7cm,width=4cm,angle=270]{fig1b.eps}
\caption{\small{Temperature {\em (upper panel)} and mass {\em (lower panel)}
profiles of NGC\thin 720. The mass data are shown with models comprising 
simple NFW (dashed line), NFW plus stellar (dotted line) and compressed NFW plus stellar 
(solid line) potentials.}} \label{figure1}
\end{figure}
\begin{table*}
\centering
\caption{\small{Best-fit values of \mvir, in units of $10^{12}$\msun, and c 
for the different mass models fitted.
Where no error is quoted the parameter value was fixed. Error bars are at 1-$\sigma$. 
Figures in square parentheses are systematic error
estimates (see text). Other figures in parentheses represent the change in
the best-fit value if \mstars/\lb\ is varied by $\pm$20\%.}}
\begin{small}
\begin{tabular}{lrllrllr} \hline
Galaxy &  \multicolumn{3}{c}{NFW} & \multicolumn{4}{c}{Compressed NFW+ stars}  \\
 & $\chi^2$/dof & \mvir & c  
& $\chi^2$/dof & \mvir & c & \mstars/\lb \\ \hline
NGC\thin 720  & 2/9 & $3.4^{+1.5}_{-0.9}$ & 47$\pm$15 & 1/9 & $6.1^{+5.2}_{-2.4}$ & 19$^{+8}_{-6}$ & 3.3 \\
& & [$^{+1.7}_{-0.7}$] & [$\pm$8] & & [$^{+15}_{-1.2}$]($^{+1.4}_{-0.9}$)& [$^{+2}_{-7}$]($\pm$4) & \\
NGC\thin 1407 & 23/11 &  $9.0^{+6.3}_{-3.5}$ & 36$^{+13}_{-9}$ & 18/11 & 300($>$30) & 4.6$\pm$4.2 & 4.7 \\
& & [$^{+4.3}_{-3.7}$] & [$^{+14}_{-8}$] & & [-250](-230) & [$^{+7.2}_{-1.9}$]($\pm$4.1) \\
NGC\thin 4125 & 23/11 & $1.0^{+0.2}_{-0.1}$ & 88$\pm$14 &  19/11 & 1.8$^{+0.8}_{-0.4}$ & 25$^{+8}_{-6}$ & 2.4  \\
& & [$^{+0.3}_{-0.4}$] & [$^{+44}_{-7}$] & & [$^{+0.2}_{-1.1}$]($^{+0.7}_{-0.3}$)& [+38]($\pm$10) &\\
NGC\thin 4261 & 23/12 & ${1.5^{+0.3}_{-0.2}}$ & 160$\pm$20 & 21/12 &  $2.6^{+1.8}_{-1.0}$ & 38$^{+23}_{-14}$ & 4.6 \\
& & [$^{+0.4}_{-0.2}$] & [$^{+10}_{-30}$] & & [$\pm$1.2]($^{+2.0}_{-0.7}$) & [$^{+23}_{-18}$]($\pm$26)\\
NGC\thin 4472 & 53/21 & $10^{+4}_{-3}$ &  30$^{+7}_{-5}$ & 30/20 & $55^{+160}_{-28}$ & 11$\pm$4 & 0.87$\pm$0.14 \\
 & & [$^{+0.4}_{-0.6}$] & [$^{+20}_{-2}$] & & [$^{+2}_{-30}$] & [$^{+1.7}_{-0.8}$]\\
NGC\thin 4649 & 30/7 & $2.5^{+0.4}_{-0.3}$ & 140$\pm$10 & 21/7 & $17^{+36}_{-9}$ & 24$\pm8$ & 4.7\\
& & [$^{+0.1}_{-1.0}$] & [$^{+30}_{-4}$] & & [$^{+2}_{-11}$]($^{+130}_{-10}$)& [$^{+13}_{-1}$]($\pm$18)\\
NGC\thin 6482 & 0.6/5& $2.3^{+0.4}_{-0.3}$ & 99$\pm$16 & 0.4/5 & $3.5^{+1.3}_{-0.9}$ & $36^{+9}_{-7}$ & 1.2\\
 & & [$^{+0.2}_{-0.1}$] & [$\pm$4] & & [$\pm$0.3]($^{+0.6}_{-0.4}$) & [$\pm$2]($\pm$9)\\\hline
\end{tabular}
\end{small}
 \label{table1}
\end{table*}

\section{Data analysis}
The \chandra\ data were processed with \ciao\ 3.2.2, following standard 
procedures. Due to the low surface brightness of the data, special care
was taken in treating the background, for which we adopted a modelling 
procedure (see Humphrey et al, 2005).
We fitted the spectra from concentric annuli with an APEC model (plus unresolved
point-source component in \dtwentyfive) to determine temperature and density.
The best-fitting abundances were similar to other early-type galaxies
\citep{humphrey05a}.

\section{Mass profiles}
The gravitating mass profiles were inferred from the temperature
and density profiles in two ways. First, we used parameterised
models for each, although we did not find a universal profile 
fitted either, and derived mass profiles under the assumption of 
hydrostatic equilibrium (we discuss the possible impact of 
low-significance asymmetries in
some systems--- \eg\ \citealt{randall04a}--- in Humphrey et al, 
2005, in prep).
The mass profiles were clearly more extended 
than the optical light, indicating significant DM. 
Within \reff, we found M/\lb\ for the gravitating matter 
varied from 2.3--9.3 \msun/\lsun.
In Fig.~\ref{figure1}, we show the best-fit temperature and 
mass profiles for NGC\thin 720.
Alternatively, we also used the temperature
profile, and an assumed mass profile (see below) to derive a 
density model, which we fitted to the data. This procedure 
gave more robust mass constraints. These techniques are 
outlined in Humphrey et al (2005).

Simple NFW fits to the data gave very large ($\gg 20$) values for
c, the halo concentration,
in contrast to the typical values predicted by simulations 
\citep[$\sim$15 \eg][]{bullock01a}. 
To investigate whether baryonic matter affects the mass profile, 
we included an
\citet{hern90} mass component to trace the stars and allowed 
the DM halo to be compressed due to baryonic condensation
\citep{gnedin04a}. Assuming all mass within \reff\ is stellar
did not give meaningful fits. Fixing stellar mass (\mstars) within 
\reff\ to be half of the total reduced c, bringing \mvir\ and 
c into better agreement with simulations. This model fitted all
the galaxies well.
We note that if adiabatic compression of the DM halo was turned
off, for a fixed \mstars/\lb, c was significantly higher.

Our results were very sensitive to \mstars/\lb, which could
only be constrained in NGC\thin 4472 (in which it was $\sim$1).
In general, though, we found  \mvir\ and c were consistent with 
simulations \citep{bullock01a}, albeit very uncertain.
In Table~\ref{table1} we show a summary of our results and, in addition,
the sensitivity to \mstars/\lb\ and the spectral analysis 
choices (\eg\ \nh\ or background modelling).

\section*{Acknowledgments}
We thank Oleg Gnedin for making available his adiabatic
contraction code. Support for this work was provided by NASA under grant
NNG04GE76G issued through the Office of Space Sciences Long-Term
Space Astrophysics Program.
\bibliographystyle{apj_hyper}
\bibliography{paper_bibliography.bib}
\end{document}